\documentclass[11pt]{article}

\usepackage[a4paper,top=2.0cm,bottom=2.0cm,left=2.5cm,right=2.5cm,nohead]{geometry}
\usepackage[english]{babel}     
\usepackage[T1]{fontenc}       
\usepackage[ansinew]{inputenc} 

\usepackage{amsmath,amsthm}
\usepackage{amssymb}
\usepackage{url}
\usepackage[numbers]{natbib}
\usepackage{graphicx,dblfloatfix}
\usepackage{float}
\usepackage{tikz}
\usetikzlibrary{calc,shapes.multipart,backgrounds,snakes}
 \tikzstyle{frame}=[framed,background rectangle/.style={draw, rounded corners}]
\usepackage[nofillcomment,vlined,linesnumbered,titlenumbered,algo2e,ruled]{algorithm2e}
\usepackage{afterpage} 
\usepackage{booktabs}

\theoremstyle{plain}

\providecommand{\myfloor}[1]{\left \lfloor #1 \right \rfloor }

\title{\bf A GPU-based parallel algorithm for enumerating all chordless cycles in graphs}

\author{Walid A. R. Jradi\and
        Elis\^angela S. Dias\and
        Diane Castonguay\and
        Humberto Longo\and
        Hugo A. D. do Nascimento\\[10pt]
        {\bf Universidade Federal de Goi\'as}\\
        {\small\tt \{walid.jradi,elisangela,diane,longo,hadn\}@inf.ufg.br}}

\begin{document}

\maketitle

\begin{abstract}
\noindent In a finite undirected simple graph, a {\it chordless cycle} is an induced subgraph which is a cycle. We propose a GPU parallel algorithm for enumerating all chordless cycles of such a graph. The algorithm, implemented in OpenCL, is based on a previous sequential algorithm developed by the current authors for the same problem. It uses a more compact data structure for solution representation which is suitable for the memory-size limitation of a GPU. Moreover, for graphs with a sufficiently large amount of chordless cycles, the algorithm presents a significant improvement in execution time that outperforms the sequential method.\\[5pt]

\noindent {\bf Keywords:}
Graphs, Chordless Cycles, Parallel Algorithm, GPU, OpenCL.
\end{abstract}

\section{Introduction}
\label{sec:introduction}

Consider a finite undirected simple graph $G=(V,E)$, with $n=|V|$ and $m=|E|$. A {\it chordless cycle} is an induced subgraph that is a cycle, i.e., there is no edge outside the cycle connecting two vertices of it.

Sequential and parallel algorithms to the problem of determining if a graph contains a chordless cycle with $k \geq 4$ vertices, for some fixed cycle length $k$, were proposed by Chandrasekharan et al.~\cite{CLM1993}. They presented an algorithm, where a cycle $C_l$, $l \geq k$ can be found in $\mathcal{O}(m^2 \cdot n^{k-4})$ time sequentially and in $\mathcal{O}(\log n)$ time using $\mathcal{O}(m^2 \cdot n^{k-4})$ processors in parallel on a CRCW PRAM. However, finding just one cycle of length greater or equal to a fixed value $k$ is easier than enumerating all chordless cycles in a graph $G$.

In general, enumeration is classified as belonging to the class of $\mathcal{P}$-complete problems, whose resolution is comparatively as hard as the resolution of problems in $\mathcal{NP}$-complete class~\citep{BM1976,V1979}. Although there are sequential algorithms to solve problems in such class, they become impractical as the problem size grows, preventing its utilization and requiring the usage of other approaches, like heuristics and meta-heuristics (trying to find a good enough approximate solution) or parallel computing (aiming the reduction of the algorithm's execution time).

Many sequential algorithms have been proposed for enumerating graph structures such as cycles~\cite{FGMPRS2013,DK1995,LW2006,LT1982,RT1975,SS2007,W2008}, circuits~\cite{B2010,T1973}, paths~\cite{HH2006,RT1975}, trees~\cite{KR2000,RT1975} and cliques~\cite{MU2004,TTT2006}. This kind of tasks is usually hard to deal with, since even a small graph may contain a huge number of such structures. Nevertheless, enumeration is necessary in the resolution process of many practical problems. In particular, the enumeration of chordless cycles is useful in areas like the study of ecological networks with the aim of discovering the predators that compete for the same prey~\cite{SBBHN2013}. Usually, a directed {\it food web} graph is transformed into a {\it niche overlap} graph to represent the competition between species, see~\cite{WW1990}. The lack of chordless cycles in the later graph means that the species can be rearranged along a single hierarchy. Another application of enumeration of chordless cycles is the prediction of nuclear magnetic resonance chemical shift values~\cite{Satoh20057431}. 

A sequential algorithm that enumerates all chordless cycles is described by Sokhn et al.~\cite{SBBHN2013}. The general principle of this algorithm is to use a vertex ordering and to expand paths from each vertex using a depth-first search (DFS) strategy. This approach has the disadvantage of finding twice each chordless cycle. Unfortunately, the authors did not present a complexity analysis of the algorithm.

Another sequential algorithm to enumerate chordless cycles was proposed by Uno and Satoh~\cite{US2014} and, as the algorithm of Sokhn et al.~\cite{SBBHN2013}, each chordless cycle will appears more than once in the output. Actually, each cycle will appear as many times as its length. Thus, the algorithm has $\mathcal{O}(n \cdot (n + m))$ time complexity in size of the sum of lengths of all the chordless cycles in the graph.

Dias et al.~\cite{DCLJ2014} developed, up to our knowledge, the fastest sequential algorithm to enumerate all chordless cycles in undirected graphs since it finds all cycles just once. It is recursive and based on a depth-first search (DFS) strategy, with $\mathcal{O}(n + m)$ time complexity in the output size,  Although the technique presented in~\cite{DCLJ2014} surpasses all solutions currently available, it still takes a considerable processing time when applied to some complex graphs or to graphs with a large amount of chordless cycles. 

Again, up to our knowledge, there is no parallel algorithm for the problem of enumerating all chordless cycles in an undirected graph. In this paper, we fill in this gap by presenting a GPU-based parallel algorithm to such problem that is fast when applied to complex graphs and/or large amount of chordless cycles.

The remaining of this paper is organized as follows. In Section~\ref{sec:background}, we establish some preliminary definitions. In Section~\ref{sec:sequential}, we present the idea of our sequential algorithm. The parallel algorithm is introduced in Section~\ref{sec:parallel}. Section~\ref{sec:results} describes the experimental tests and the results produced by the new algorithm compared againts other methods. Conclusion and future work are discussed in Section~\ref{sec:conclusions}.

\section{Background}
\label{sec:background}

In this section, we present some mathematical definitions that support our approach to enumerate all chordless cycles of a graph. For more details on these definitions, see \citep{DCLJ2014}.

Let $G = (V, E)$ be a finite undirected simple graph with vertex set $V$ and edge set $E$. Let $n=|V|$ and $m=|E|$. We denote by $Adj(x) = \{y \in V \>|\> (x, y) \in E\}$ the set of neighbors of a vertex $x \in V$ and by $Adj[x] = \{x\} \cup Adj(x)$ the closed neighborhood of $x$.

A \textit{simple path} is a finite sequence of vertices $\langle v_1, v_2, \dots, v_k \rangle$ such that $(v_i, v_{i+1})\allowbreak \in E$ and no vertex appears repeated in the sequence, that is,  $v_i\neq v_j$, for $i,j \in \{1, \dots, k-1\}$ and $i \neq j$. A \textit{cycle} is a simple path $\langle v_1, v_2, \dots, v_k \rangle$ such that $(v_k, v_1)\in E$. We denote a cycle with $k$ vertices by $C_k$\footnote{Note that our definition of cycle does not repeat the first vertex at the end of the sequence as usually done by other authors. We decided to use this definition (with the first vertex implicitly included at the end) because it simplifies the representation of a rotated version of the cycles. If $\langle v_1, v_2, \dots, v_k \rangle$ is a cycle, so also are $\langle v_i, v_{i + 1} \dots, v_k, v_1, v_2, \dots, v_{i-1} \rangle$ and $\langle v_i, v_{i - 1}, \dots, v_2, v_1, v_k, \dots, v_{i+1} \rangle$, for all $i = 1, \dots, k$.}. A {\it chord} of a path (resp. cycle) is an edge between two vertices of the path (cycle), that is not part of it. A path (cycle) without chord is called a {\it chordless path (chordless cycle)}. 

The minimum degree among all vertices of $G$ is denoted by $\delta(G)$. The maximum degree is denoted by $\Delta(G)$; for reason of simplicity, we use just $\Delta$. We represent by $d_{G}(v)$ the degree of a particular vertex $v\in V$. The subgraph induced by the subset $V - X$, for $X \subseteq V$ ($V - \{u\}$, for $u \in V$), is denoted by $G - X$ ($G - u$). 

An ordering of the vertices of $G$ can be defined by a bijection $\ell: V \to \{1, 2, \dots, n\}$.  We call such a bijection a {\it vertex labeling}. Given a such vertex labeling, if
 \begin{enumerate}
  \item \label{c-a} $G$ contains a simple cycle $\langle v_1, v_2, \dots, v_k \rangle$,
  \item \label{c-b} $\ell(v_2) =\allowbreak \min\{\ell(v_i) \>|\> i= 1, \dots, k\}$ and
  \item \label{c-c} $\ell(v_1) < \ell(v_3)$,
 \end{enumerate}
then $\ell$ defines the cycle in a unique way. Note that any cycle can be described as $\langle v_i, \dots,\allowbreak v_k, \allowbreak v_1, v_2, \dots, v_{i-1} \rangle$ or $\langle v_i, \allowbreak \dots, v_2, v_1, v_k, \dots, \allowbreak v_{i+1} \rangle$, for all $i = 1, \dots, k$. Let $i$ be a vertex index such that $\ell(v_{i}) = \min\{\ell(v_j)$ $|\> j=1, \dots, k\}$. There are only two possibilities for the vertex $v_i$ to be the second one of the cycle: $\langle v_{i-1},v_{i},v_{i+1}, \dots, v_k, v_1, v_2, \dots, v_{i-2} \rangle\>$ or $\>\langle v_{i+1},v_{i},v_{i-1}, \dots, v_2, v_1, v_k, \dots, \allowbreak v_{i+2} \rangle$. Since the neighbors of $v_i$ in the cycle are $v_{i-1}$ and $v_{i+1}$, exactly one of these possibilities satisfies the condition~\ref{c-c}.

In the approach introduced in~\cite{DCLJ2014}, a vertex labeling is given by a particular bijection $\ell: V(G) \to \{1, \ldots, n\}$ called {\it degree labeling}. It is constructed over a sequence of subgraphs of $G$, starting with $G_1 = G$. For $i \geq 1$, the $(i+1)^{\mbox{th}}$ subgraph is defined as $G_{i+1} = G_{i} - u_{i}$, for a chosen $u_i \in V(G_i)$ such that $d_{G_{i}}(u_{i}) = \delta(G_{i})$. Given such a sequence, the degree labeling is defined as $\ell(u_i) = i$ for each $i$. 

A {\it triplet} is defined as a sequence of vertices that can initiate a chordless path of length greater than three. Let $T(G)$ denote the set of all initial valid triplets of $G$, that is, $T(G) = \{ \langle x, u, y \rangle \mid x, u, y \in V \mbox{ with } x, y \in Adj(u)$, $\ell(u) < \ell(x) < \ell(y)$ and $(x, y) \notin E\}$. The above labeling scheme allows to find every chordless cycle only once and to begin with a smaller initial set of chordless paths, which reduce considerably the search space. Note that, for any chosen degree labeling, if $G$ is a tree then there are no possible triplets, that is, $T(G) = \varnothing$. Moreover, if $G$ has a unique cycle then $|T(G)| = 1$, no matter what degree labeling is used, that is, unneeded triplets are discarded. As detailed in~\cite{DCLJ2014}, an upper bound for the initial search space size is given by $|T(G)| \leq \frac{(\Delta - 1) \cdot m}{2}$.

Given a chordless path $p = \left \langle v_1, v_2, \ldots, v_k \right \rangle$ and a vertex $v \in Adj(v_k)$ such that $v \neq v_{k-1}$, thus exactly one of the following occurs:
 \begin{enumerate}
   \item \label{p-a}  $\left \langle p, v \right \rangle=\langle v_1, v_2, \ldots, v_k, v \rangle$ is a chordless path;
   \item \label{p-c}  there exists $i\in\{1, \ldots, k-1\}$ such that $p = \left \langle v_i, v_{i+1}, \ldots, v_k, v \right \rangle$ is a chordless cycle.
 \end{enumerate}
 
 Since $v \in Adj(v_k)$, $v \neq v_{k-1}$ and $p$ is a chordless path, then $\langle p, v\rangle$ is a simple path. Suppose that $\langle p, v\rangle$ is not a chordless path. Therefore, there is an index $i \in \{1, \ldots, k-1\}$ with $(v, v_i) \in E$. Choosing the biggest index $i$ with this property, we have the desired chordless cycle. Case~\ref{p-a} states that path $\langle p, v\rangle$ is an expansible chordless path. Case~\ref{p-c}, with $i \neq 1$, state that path $\langle p, v\rangle$ has a chord or, with $i=1$, $\langle p, v\rangle$, is a chordless cycle.

\section{The sequential algorithm}
\label{sec:sequential}

The sequential algorithm for chordless cycles enumeration of Dias et al.~\cite{DCLJ2014} is briefly described here in order to help the understanding of the proposed, parallel approach. Further details and experimental results can be found in~\cite{DCLJ2014}. A pseudo-code of this algorithm is presented in Algorithm~\ref{alg:cicloSemCordaAltoNivel}.

A degree labeling is initially calculated for the input graph $G$ (Line~\ref{cc-step-0}). Then, the set of initial valid triplets $T(G)$ (Line~\ref{cc-step-1}) is computed, as described previously in Section~\ref{sec:background}. The set $C$ is initialized (Line~\ref{cc-step-2}) with all triangles (which are also chordless) and variable $T$ receives the set $T(G)$ (Line~\ref{cc-step-3}). Next, for each triplet $t=\langle x,u,y\rangle \in T$, a DFS strategy is used to check the existence of a chordless cycle starting at its last vertex ($y$) and respecting the constraints on the labeling order. Line~\ref{cc-step-8} of the algorithm verifies if the addition of a neighbor of $y$ to the path gives:
 \begin{enumerate}
   \item\label{case1} a chordless cycle;
   \item\label{case2} a chord in the current path; or
   \item\label{case3} another expansible path.
 \end{enumerate}

In case~\ref{case1}, the newly chordless cycle found is added to the set $C$ of cycles (Line~\ref{cc-step-10}); in case~\ref{case2}, the path is discarded and, in the last case, the expanded path is added to the set $T$ of expandable paths (Line~\ref{cc-step-11}). The same process is repeated until the set $T$ becames empty.

Due to the initial conditions of the triplets and the way the search is performed, the algorithm finds all chordless cycles and still avoids rotations of the same solution (two or more cycles with the same structure but that start at different vertices). This provides a faster execution of the algorithm. Dias et al.~\citep{DCLJ2014} presented another version of the algorithm, that uses a specialized breadth-first search (BFS), to ensure that each path expansion finds a chordless cycle. However, in practice, the algorithm without BFS leads to a shorter execution time.

\begin{center}
\begin{minipage}{.95\textwidth}
\begin{algorithm2e}[H]
   \SetKwFunction{ccvisit}{CC\_Visit}
   \SetKwFunction{dl}{\it DegreeLabeling}
   \LinesNumbered

   \KwIn{Graph $G$.}
   \KwOut{Set $C$ of all chordless cycles of $G$.}
   \BlankLine

   {\bf perform} {\it DegreeLabeling(G)}\nllabel{cc-step-0}

   $T(G) \leftarrow \{ \langle x, u, y \rangle \mid x, u, y \in V: x, y \in Adj(u); \ell(u) < \ell(x) < \ell(y)$ and $(x, y) \notin E\}$\nllabel{cc-step-1}
   \BlankLine

   $C \leftarrow \{ \langle x, u, y \rangle \mid x, u, y \in V: x, y \in Adj(u); \ell(u) < \ell(x) < \ell(y)$ and $(x, y) \in E\}$\nllabel{cc-step-2}\\
   $T \leftarrow T(G)$\nllabel{cc-step-3}
   \BlankLine

   \While{$(T \neq \varnothing)$}{\nllabel{cc-step-4}
      $p \leftarrow \langle v_1, v_2, \ldots, v_t \rangle \in T$\nllabel{cc-step-5}\\
      $T \leftarrow T - \{p\}$\nllabel{cc-step-6}
   \BlankLine

   \ForEach{$v \in Adj(v_t)$}{\nllabel{cc-step-7}
	\If{$((\ell(v) > \ell(v_2))\ \mathbf{and}\ (v \notin Adj(v_i),\> i \in \{2, \dots, t-1\}))$\nllabel{cc-step-8}}{
	  \eIf{$v \in Adj(v_1)$\nllabel{cc-step-9}}{
	    $C \leftarrow C \cup \{\langle p, v \rangle\}$\nllabel{cc-step-10}
	  }
	  {
	    $T \leftarrow T \cup \{\langle p, v \rangle\}$\nllabel{cc-step-11}
	  }
	}
       }
      \BlankLine
    }
   \BlankLine
   \Return{$C$.}

   \caption{\textit{SequentialChordlessCycles(G)}
   \label{alg:cicloSemCordaAltoNivel}}
\end{algorithm2e}
\end{minipage}
\end{center}


A possible strategy for the parallelization of Algorithm~\ref{alg:cicloSemCordaAltoNivel} is the extension of multiple chordless paths through the simultaneous checking of the feasibility of adding each one of the neighbors of the last vertex on each path. The following sections detail this approach.

\section{A GPU-based parallel algorithm}
\label{sec:parallel}

In this section, we present our parallel approach for the chordless cycles enumeration problem. The strategy adopted for parallelizing the computation done by Algorithm~\ref{alg:cicloSemCordaAltoNivel} is to split it into two stages and define a parallel approach for each one of them. The first stage involves the creation of the set $C$, with all cycles of length three, and the set $T(G)$, with all initial valid triplets $\langle x, u, y\rangle$ (Lines \ref{cc-step-1}--\ref{cc-step-3} of Algorithm~\ref{alg:cicloSemCordaAltoNivel}). The second stage gets each path $\langle x, u, y, \ldots, v\rangle$ in $T(G)$, that characterizes a chordless path, and tries to extend it by adding a neighbor to the last vertex, $v$ (Lines~\ref{cc-step-4}--\ref{cc-step-11}). 

The computation of a degree labeling (that appears at Line~\ref{cc-step-0} of Algorithm~\ref{alg:cicloSemCordaAltoNivel}), however, was not parallelized. Due to its inherent sequential nature and to the low impact in the processing time of the algorithm, this step was kept sequential as a preprocessing task and the resultant labels were used in our parallel stages. 

The final parallel algorithm was mapped to a GPU architecture, which basic concepts are presented just below.

\subsection{GPU Programming}
\label{sec:GPU_programming}

General-Purpose programming on Graphics Processing Units (GP-GPU) technique consists in the utilization of GPUs to run non-graphical applications. GPUs are \textit{stream processors} -- that is, execution units able to operate in parallel, simultaneously performing routines in a large amount of data. They are focused on data parallelism and the basic idea is to maximize the data flow rate rather than to minimize latency (as in CPUs). GPU architecture emphasizes the implementation of many ``light'' threads concurrently, instead of executing a few traditional heavy threads.

In this work, it is adopted a very common GPU architecture consisting of \textit{p} Symmetric Multiprocessors (SM) and a large, but slow, global memory. These processors, in turn, are grouped into larger units called \textit{Symmetric Multiprocessors} (SM). Each SM, therefore, has a subset of $\frac{p}{|SM|}$ processors, where $|SM|$ is the number of SM units. All processors within a SM can communicate through a small, but fast, local shared memory. Processors in different SMs can only communicate through the global memory. There is also a private memory, unique to each running thread, which is much faster but smaller than the other memories.

Developing efficient data structures using the GPU memory model is a challenging task by itself~\cite{lefohn2006glift}. The distinct characteristics of GPU and CPU architectures make many ordinary data structures (as the ones used in the sequential algorithm described in~\citep{DCLJ2014}) not suitable for parallelization in GPUs. Thus, other data structures and data flows have to be employed to overcome limitations (such as small memory size) and to take advantages of the characteristics of the different types of the GPU memories.

Next, problems with the data structures of the sequential algorithm are discussed and new data structures for the parallel algorithm are presented. After that, the following section details the two parallel stages mentioned above.

\subsection{Data structures}

Usually, graphs are represented by adjacency matrices or lists. Although an adjacency matrix allows verifying connectivity between two vertices in constant time, it has three primary issues:

\begin{itemize}
  \item for sparse graphs, there is a significant waste of memory space;
  \item due to the large space occupied, it is not possible to allocate the entire matrix in the fast, but small, GPU local shared memory. Even in advanced GPU models, this memory does not exceed 64KB. A simple graph containing just 256 vertices would be enough to fill in all this memory ($256 \cdot 256 \cdot 1$ byte $= 65536$ bytes) with such a data structure;
  \item the storage in the GPU global memory leads to severe degradation performance, because its access time is much larger than that of the local memory of each set of SMs (Symmetric Multiprocessors).
\end{itemize}

Consequently, the use of adjacency lists, which allow a more compact graph representation, would be justified. However, its variable size for each vertex list does not allow an efficient implementation on GPUs.

To overcome such problems, we used an adapted version of the compact graph representation proposed by Harish and Narayanan~\cite{HN2007}. Our version of this representation is composed by three vectors, $V_e$, $E_e$ and $L_v$. Vector $V_e$ stores vertices of a graph $G = (V, E)$, in a way that a vector index is the original vertex identification and the corresponding vector content indicates the position of its first neighbour in the adjacency vector $E_e$. Since the graph is undirected, it is necessary to represent each edge $(i, j) \in E$ in both adjacent lists of $i$ and $j$. So, $|E_e| = 2 \cdot |E|$. Vector $L_v$ stores the degree labels associated to each vertex of the graph. If the lists of adjacent vertices are kept sorted in $E_e$, then the check whether two vertices are adjacent can be performed in time $\mathcal{O}(\log \Delta)$ by a binary seach. 

\begin{figure}[hbt]
\tiny
 \centering
 \includegraphics[scale=.9]{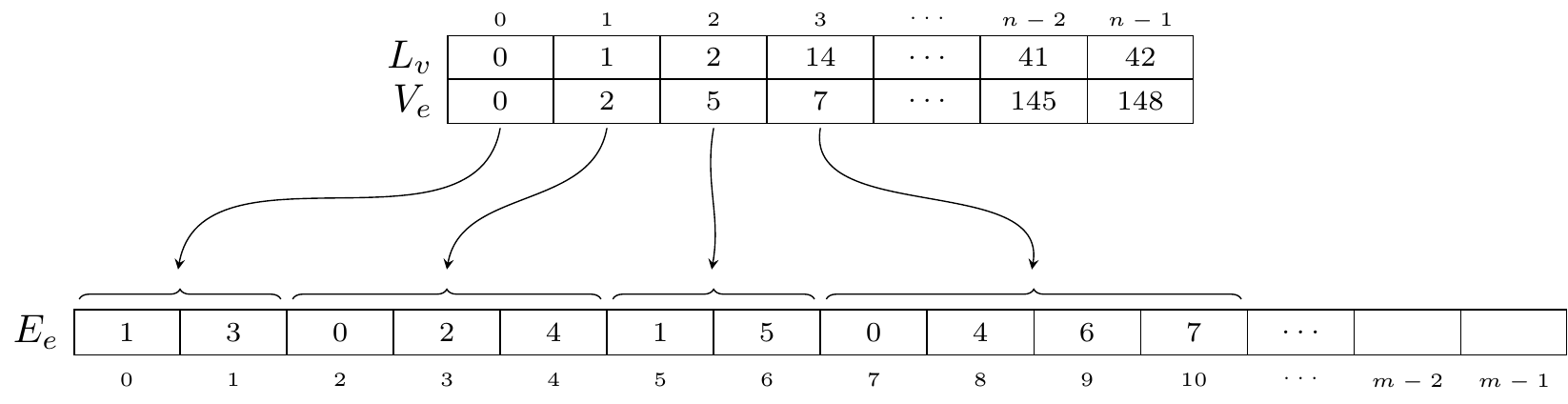}
 \caption{Compact representation of a graph.}
 \label{fig:Compact_Representation}
\end{figure}

Figure~\ref{fig:Compact_Representation} illustrates this compact representation, where vertex 0 is neighbor of vertices 1 and 3, vertex 1 is neighbor of 0, 2 and 4 and so on. Considering 2 bytes for an adjacency index, this representation takes only $(|V| + |E|) \cdot 2 \cdot 2$ bytes. This is small enough to store a graph of size at most 32KB in the fast local/shared memory of each SM in the majority of GPUs currently available. The search time for finding the neighbors of a vertex in this data structure is $\mathcal{O}(\Delta)$, even for dense graphs.

To allow more efficient storage of partial and complete solutions (chordless paths and chordless cycles, respectively), a map of bits was employed. A single bit is enough to indicate whether a vertex belongs to a solution because it is not important to store the vertices order in the chordless paths/cycles. This map is defined by a bi-dimensional matrix $S$ that contains a row for each chordless path/cycle and $n$ columns of bits, one for each vertex of the graph. In terms of bytes, the number of columns is $\lceil \frac{n}{8}\rceil$. Vertex $v_j$ belongs to path/cycle $i$ if, and only if, bit $j$ of row $i$ is 1. Despite the fact that such bitmap does not provide a vertices visit order, it depicts unambiguously each chordless path/cycle in the graph $G$.

In addition to the small occupied space, this data structure allows to add a vertex to a solution by a simple bitwise operation. Bit-level operations are among the least computationally expensive ones.

\begin{figure}[!b]
   \centering
   \includegraphics[width = 0.7\textwidth]{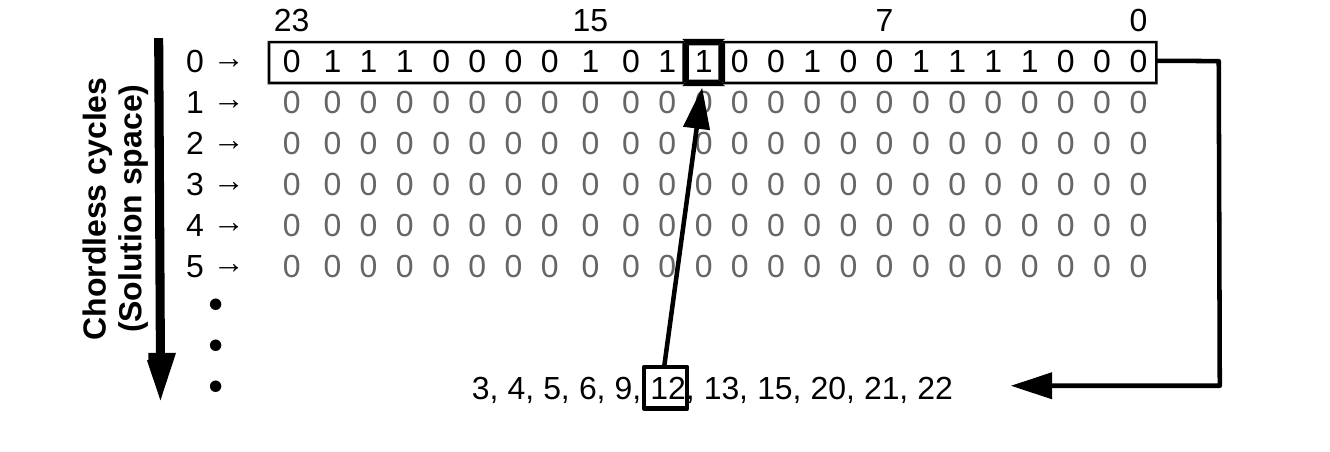}
    \caption{Solution Space, where each vertex occupies just one bit.}
   \label{fig:Solution_Space_2}
\end{figure}

Figure~\ref{fig:Solution_Space_2} shows an example of this data structure. Row 0 contains a combination of bits that describes a chordless cycle in a graph $G$ with $n \leq 24$. In this case, regardless the number of vertices in the graph, a path/cycle storage occupy only 3 bytes in the worst case.

However, with this matrix, it is not possible to know neither the latest vertex added to a chordless path, nor the initial or the second vertex of the path. These pieces of information are essential to the algorithm, as the last vertex is used for expanding the path, while the initial vertex allows to check whether the path forms a chordless cycle or not, and the second vertex of a path takes part of a labeling condition check.

To circumvent this problem, three auxiliary vectors, $V_1, V_2$ and $V_L$, are used. $V_1$ and $V_2$ stores the first and the second vertex of the paths and the content of their cells never changes once they were set. $V_L$ stores the last vertex added to the chordless paths and its content is updated whenever a path is expanded. The sizes of $S, V_1, V_2$ and $V_L$ have to be sufficiently large to contain information about all chordless paths that are being processed  at any given moment.

Once the number of rows in each vector/matrix equals the number of chordless paths, these data structures can potentially occupy a large space in memory. Thus, they are kept in the global memory of the GPU. Even further, as we describe later, in Section~\ref{sec:SecondStage}, these data structures are replicated in order to speed up the processing of chordless paths.

Now we explain the parallel implementation of our approach involving two stages.

\subsection{First Stage}
\label{sec:FistStage}

The first stage involves the parallelization of Lines~\ref{cc-step-1} to~\ref{cc-step-3} of Algorithm~\ref{alg:cicloSemCordaAltoNivel}, which computes the sets $C$ and $T(G)$. Such sets are created by selecting every vertex $u\in V$ and analyzing all pairs $x$ and  $y$ of adjacent vertices to $u$, such that $\ell(u) < \ell(x) <\ell(y)$. If $(x,y) \in E$, then the triplet $\langle x,u,y \rangle$ is a simple circle and is added to $C$; otherwise, $\langle x,u,y \rangle$ is a chordless path and is added to $T$. This process, when done sequentially, demands time $\mathcal{O}(|V| \cdot \Delta^2)$. 

Our parallel approach for this stage is described in Algorithm~\ref{alg:find_initial_triplets_parallel}. It consists of starting $M=(|V| \cdot \Delta^2(G))$ parallel threads in the GPU. Each thread $j$  uses its unique global identifier, denoted by $gId(j)$, to compute the indices $i_x, i_u$ and $i_y$ of the vertices, respectively, of a triplet $\langle x,u,y \rangle$ in the compact graph representation (see Lines~\ref{ltparallel-step-2} to~\ref{ltparallel-step-4} of Algorithm~\ref{alg:find_initial_triplets_parallel}):
\begin{align}
 i_u \leftarrow & \myfloor{\frac{gId(j)}{\Delta^2}};\\
 i_x \leftarrow & \myfloor{\frac{gId(j) - i_u \cdot \Delta^2}{\Delta^2}};\\
 i_y \leftarrow & gId(j) \bmod \Delta;
\end{align}
where $i_u$ is the index of vertex $u$ in the vector $V_e$; $i_x$ and $i_y$ are relative indices of $x$ and $y$ in the vector $E_e$. Index $i_u$ ranges from $0$ to $|V|-1$, and $i_x$ and $i_y$ ranges from 0 to $\Delta -1$. 

Values $i_x$ and $i_y$ are used to determine two neighbors of the vertex $u$. They have to be added to the value $V_e[i_u]$ in order to obtain absolute indices in $E_e$. However, $i_x$ and $i_y$ should only be employed if they refer to valid neighbors (that is, if they are less or equal to the amount of adjacent vertices of $u$). Such analysis is carried out in Algorithm~\ref{alg:find_initial_triplets_parallel} by Lines~\ref{ltparallel-step-5} to~\ref{ltparallel-step-10}. The functions $neighborsLowerBound(u)$ and $neighborsUpperBound(u)$ return, respectively, the absolute indices of the first and of the last neighbors of $u$ in $E_e$, what allows to validate the indices $i_x$ and $i_y$ (Lines~\ref{cc-step-7}--\ref{cc-step-8}).
  
Finally, with valid vertices $u$, $x$ and $y$, each thread tests the label condition $\ell(u) < \ell(x) <\ell(y)$. Only the threads for which this label condition is satisfied continue their execution. They check whether $x$ is a neighbor of $y$ and, if true, the triplet $\langle x,u,y\rangle$ is added to a set $C$ of chordless cycles. Otherwise, the triplet is added to the set $T(G)$ of initial valid triplets (chordless paths). 
 
\begin{figure}[!b]
  \centering
  \includegraphics[scale=.9]{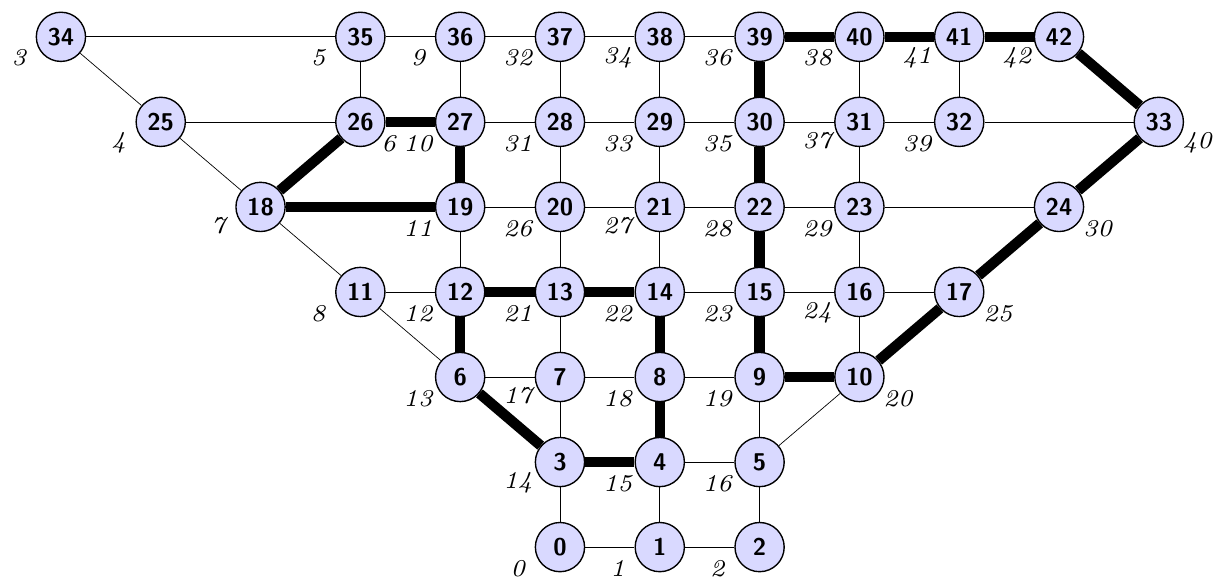}
  \caption{Part of the downtown area of the city of Goi\^ania, Goi\'as, Brazil, with degree labels near to each vertex. In highlight, three of the 9316 possible chordless cycles for the graph.}
  \label{fig:Goiania_Downtown}
\end{figure}

As an example of the algorithm, and referring to the compact graph representation in Figure~\ref{fig:Compact_Representation}, to the graph in Figure~\ref{fig:Goiania_Downtown} and consider a thread $j$ with global unique identifier $gId(j)=1$. That thread sets $i_u = 0, i_x = 0$ and $i_y = 1$. Next, the thread defines $k_1=V_e[i_u]=0$, $k_2=V_e[i_u+1]-1=1$, $u=0$, $x=E_e[k_1+i_x]=1$ and $y=E_e[k_1+i_y]=3$. Using a particular labeling, e.g., $\ell(u)=0$, $\ell(x) = 1$ and $\ell(y) = 14$, the condition $\ell(x) < \ell(u) < \ell(y)$ is satisfied and $\langle x,u,y\rangle$ is an initial valid triplet. Then, since $x$ and $y$ are not adjacent, the triplet $\langle x, u, y\rangle$ is inserted into $T(G)$. 

Lines~\ref{ltparallel-step-2} to~\ref{ltparallel-step-12} of Algorithm~\ref{alg:find_initial_triplets_parallel} demand constant time, while Line~\ref{ltparallel-step-13} is $\mathcal{O}(\Delta)$. Lines~\ref{ltparallel-step-14} and~\ref{ltparallel-step-15} require serialization in the index calculation in order to write $\langle x, u, y\rangle$ into $C$ or $T(G)$ in the right position. In the worst case, $\mathcal{O}(|V| \cdot \Delta^2)$ threads may try to perform such writing operations simultaneously, but the experiments carried out show that this occurs a small amount of times for many large graphs. Moreover, this serialization is much faster than the computations done by the other lines of Algorithm~\ref{alg:find_initial_triplets_parallel} and it is necessary only to allocate a free memory position to write the chordless path or cycle. The writing operation, by itself, can be done in parallel.

\begin{center}
 \begin{minipage}[t]{0.9\textwidth}
  \begin{algorithm2e}[H]
   \LinesNumbered
   \BlankLine
   \KwIn{Compact representation of an undirected simple graph $G = (V,E)$.}
   \KwOut{Set $T(G)$ of initial chordless paths of length 3.}
   \BlankLine

    \ForPar{each thread $j$, with $j = 0, \dots, |V| \cdot \Delta^{2} - 1$\nllabel{ltparallel-step-1}}
    {
     $i_u \leftarrow \myfloor{\frac{gId(j)}{\Delta^{2}}}$\nllabel{ltparallel-step-2}\\
     $i_x \leftarrow \myfloor{\frac{gId(j) - i_u \cdot \Delta^{2}}{\Delta}}$\nllabel{ltparallel-step-3}\\
     $i_y \leftarrow gId(j) \bmod \Delta$\nllabel{ltparallel-step-4}\\

     $k_1 \leftarrow neighborsLowerBound(u)$\nllabel{ltparallel-step-5}\\
     $k_2 \leftarrow neighborsUpperBound(u)$\nllabel{ltparallel-step-6}\\

     $u \leftarrow i_u$\nllabel{ltparallel-step-7}\\
     $x \leftarrow (-1) \cdot (i_x>(k_2-k_1))+(E_e[k_1+i_x]) \cdot (i_x \leq (k_2-k_1))$\nllabel{ltparallel-step-8}\\
     $y \leftarrow (-1) \cdot (i_y>(k_2-k_1))+(E_e[k_1+i_y]) \cdot (i_y \leq (k_2-k_1))$\nllabel{ltparallel-step-9}\\

     \If(\tcc*[h]{both vertices must be valid}){$((x \neq -1)$ {\bf and} $(y \neq -1))$\nllabel{ltparallel-step-10}}
     {
      $\ell(x) \leftarrow L_v(x); \> \ell(u) \leftarrow L_v(u); \> \ell(y) \leftarrow L_v(y)$\nllabel{ltparallel-step-11} \\
      \If(){$((\ell(u) < \ell(x))$ {\bf and} $(\ell(x) < \ell(y)))$\nllabel{ltparallel-step-12}}
      {
       \eIf(){$x \in Adj(y)$\nllabel{ltparallel-step-13}}
       {
        $C \leftarrow C \cup \{\langle x, u, y \rangle\}$\nllabel{ltparallel-step-14}\\
       }
       {
        $T(G) \leftarrow T(G) \cup \{\langle x, u, y \rangle\}$\nllabel{ltparallel-step-15}\\
       }
      }
     }
    }
 
   \caption{$FindingInitialTripletsParallel(G)$
   \label{alg:find_initial_triplets_parallel}}
  \end{algorithm2e}
 \end{minipage}
\end{center}

\subsection{Second Stage}
\label{sec:SecondStage}

The second stage of our approach, described in Algorithm~\ref{alg:cc_visit_parallel}, parallelizes Lines~\ref{cc-step-4} to~\ref{cc-step-11} of Algorithm~\ref{alg:cicloSemCordaAltoNivel}. It uses all processors of the GPU in parallel for evaluating the possibility of expanding the chordless paths computed in Stage 1 (and saved in $T(G)$).  This is done by allocating $\Delta$ parallel threads for every chordless path $p = \langle v_1, v_2, \ldots, v_{t-1}, v_t\rangle$ in $T(G)$. Each one of the $\Delta$ threads analyzes a potential neighbor $v$ of $v_t$.

Lines~\ref{cc-visit-parallel-step-5} and~\ref{cc-visit-parallel-step-6} of Algorithm~\ref{alg:cc_visit_parallel} defines which chordless path $p$ will be processed by thread $j$. Lines~\ref{cc-visit-parallel-step-7} to~\ref{cc-visit-parallel-step-10} specifies the neighbor $v$ of $v_t$. If $v_t$ has less than $\Delta$ neighbors, then there will be some exceeding threads. Such threads will fall in the condition $v = -1$, in Line~\ref{cc-visit-parallel-step-11}, and they will do nothing. Finally, Lines~\ref{cc-visit-parallel-step-12} to~\ref{cc-visit-parallel-step-15} perform a task according to two cases that are similar to what we have at Stage 1:

\begin{enumerate}
 \item If $v$ is adjacent to $v_1$ but not to other vertices in $\langle v_2, \ldots, v_{t-1}\rangle$, then $v$ forms a cycle and $\langle p,u\rangle$ is added to $C$;  
  \item If $v$ is adjacent only to $v_t$ then $v$ expands $p$ and the new path $\langle p,u\rangle$ is saved in a new solution map $T'$;
\end{enumerate}

Some implementation details of our algorithm need to be explained. Firstly, every extended path $\langle p,u\rangle$ is added to $T'$ instead of to $T$. We do that because it is faster to build a new data structure (for holding the extended chordless paths) than having to update $T$. In the latter case, it would be necessary to remove $\langle p\rangle$ from $T$ in addition to adding $\langle p,u\rangle$ to this set. Secondly, we use the concept of \textit{Persistent Threads}~\cite{gupta2012study} to perform the work when there are more combinations of $|T|$ paths versus $\Delta$ neighbors than parallel processors. The loop at Line~\ref{cc-visit-parallel-step-4} does this job, by iterating the analysis for a new $p\in T$.

When all threads end their processing, they have to be restarted for working on the new set $T$. This task is carried out by the host process, running on the CPU, that replaces $T$ by the recently created $T'$, and launches all threads again.  Note, however, that we do not implement the stop condition in the host as a check $T'\neq \varnothing$. This would lead to constant communication between CPU and GPU, significantly degrading the performance of the algorithm. Instead,  it is preferably a simpler approach that avoids this data transfer and that has shown to be faster: to restart all threads $|V|-3$ times. This number of steps is sufficient for our aim, since every chordless path is increased with a new vertex of $V$, moved to the set $C$ or simply discarded at each iteration of the loop. Algorithm~\ref{alg:hostprocess} illustrates the host process. 

In Algorithm~\ref{alg:cc_visit_parallel}, Line~\ref{cc-visit-parallel-step-5} and Lines~\ref{cc-visit-parallel-step-7} to~\ref{cc-visit-parallel-step-11} demand constant time. Line~\ref{cc-visit-parallel-step-6} copies a chordless path from the global GPU memory to a private thread memory. Since $\frac{|V|}{8}$ bytes are necessary to store the path, we can consider that this line takes time $\mathcal{O}(|V|)$. Lines~\ref{cc-visit-parallel-step-13} and~\ref{cc-visit-parallel-step-15} have time complexity $\mathcal{O}(\Delta\cdot t)$ for a given chordless path $p$ under analysis, because it has to perform $\mathcal{O}(t)$ adjacency checks $(t\leq |V|)$, each one of them demanding $\mathcal{O}(\log \Delta)$ verifications. Lines~\ref{cc-visit-parallel-step-14} and~\ref{cc-visit-parallel-step-16} are $\mathcal{O}(1)$, but they depend implicitly on synchronized written operations on $C$ and $T$. In the worst case, $|SM|\cdot MaxSMSize$ threads would try to access one of these sets at the same time.

Therefore, the total worst-case time complexity of Algorithm~\ref{alg:cc_visit_parallel}, as a single thread execution of Line~\ref{host-process-03} of Algorithm~\ref{alg:hostprocess}, is $\frac{|T|\cdot \Delta}{|SM|\cdot MaxSMSize}\cdot(\mathcal{O}(\log \Delta\cdot t)+\mathcal{O}(|SM|\cdot MaxSMSize)) = \mathcal{O}\left(\frac{ |T|  \cdot \Delta \cdot \log \Delta \cdot t}{|SM| \cdot MaxSMSize}\right)$ + $\mathcal{O}(|T|\cdot \Delta)$. The second part of the time complexity is due to the synchronization process. 

Consequently, Algorithm~\ref{alg:hostprocess} has time complexity $\sum_{i=1}^{|V|}{\mathcal{O}\left(\frac{|T_i| \cdot \Delta \cdot \log \Delta \cdot t_i}{|SM| \cdot MaxSMSize} +  |T_i|\cdot \Delta \right)}$, where $|T_i|$ is the size of the set of chordless paths in iteration $i$ and $t_i=\Theta(i)$. Despite such a complexity seems high, the hidden constant for the synchronization step is very low and many threads fall in the case where neither $C$ nor $T$ are updated. Another aspect to note is that $|T_i|$ is not necessarily the same over all iterations of the loop \verb|for| in Algorithm~\ref{alg:hostprocess}. So, the amount of computation performed can vary in each iteration. This will be illustrated later, in Section~\ref{sec:results}.

Regarding the space complexity, it is not possible to make a prediction about the amount of space that will be used. Depending on the structure of the graph under analysis, the amount of cordless cycles is potentially large.

\begin{center}
 \begin{minipage}[t]{0.9\textwidth}
  \begin{algorithm2e}[H]
   \LinesNumbered
   \BlankLine
   \KwIn{Compact representation of an undirected simple graph $G = (V,E)$ and list $\ell$ of labels.}
   \KwOut{Sets $T$ and $C$ of chordless paths and cycles, respectively.}
   \BlankLine

    $globalSize \leftarrow |SM| \cdot MaxSMSize$\nllabel{cc-visit-parallel-step-1}\\

    \ForPar {each thread $j$, with $j = 0, \ldots, globalSize - 1$ \nllabel{cc-visit-parallel-step-2}}
    {
     $gId(j) \leftarrow j$\nllabel{cc-visit-parallel-step-3}\\
     \While{$(gId(j) < |T| \cdot \Delta)$\nllabel{cc-visit-parallel-step-4}}
     {
      $i_p \leftarrow \myfloor{\frac{gId(j)}{\Delta}}$\nllabel{cc-visit-parallel-step-5}\\
      $p \leftarrow getCurrentPath(T, i_p)$\nllabel{cc-visit-parallel-step-6} \tcp{\footnotesize $p$ is represented here as $\langle v_1, v_2, \ldots, v_t\rangle$.}

      $k_1 \leftarrow neighborsLowerBound(v_t)$\nllabel{cc-visit-parallel-step-7}\\
      $k_2 \leftarrow neighborsUpperBound(v_t)$\nllabel{cc-visit-parallel-step-8}\\
      
      $i_v \leftarrow gId(j)\mod \Delta$\nllabel{cc-visit-parallel-step-9}\\
      $v \leftarrow -1 \cdot (i_v > (k_2-k_1))+(E[k_1+i_v]) \cdot (i_v \leq (k_2-k_1))$\nllabel{cc-visit-parallel-step-10}\\
      
      \If(){$(v \neq -1)$ {\bf and} $(v \notin p)$ {\bf and} $(L_v(v) > L_v({v_2})))$\nllabel{cc-visit-parallel-step-11}}
      {
       \If(){$(v \in Adj(v_1))$ {\bf and} $(v \notin Adj(v_i),\> i \in \{2, \dots, t-1\})$\nllabel{cc-visit-parallel-step-12}}
       {
        $C \leftarrow C \cup \{\langle p, v \rangle\}$\nllabel{cc-visit-parallel-step-13}\\
       }
       \If{$(v \notin Adj(v_i),\> i \in \{1, \dots, t-1\})$\nllabel{cc-visit-parallel-step-14}}
       {
        $T' \leftarrow T' \cup \{\langle p, v \rangle\}$\nllabel{cc-visit-parallel-step-15}\\
       }
      }
      $gId(j) \leftarrow gId(j) + globalSize$\nllabel{cc-visit-parallel-step-16}\\
     }
    }
 
   \caption{$ExpandingChordlessPathsParallel(G, \ell)$
   \label{alg:cc_visit_parallel}}
  \end{algorithm2e}
 \end{minipage}
\end{center}

\begin{center}
 \begin{minipage}[t]{0.9\textwidth}
  \begin{algorithm2e}[H]
   \LinesNumbered
   \BlankLine
   \KwIn{Compact representation of an undirected simple graph $G = (V,E)$ and a list $\ell$ of labels.}
   \KwOut{Set $C$ of chordless cycles.}
   \BlankLine

    Create the data structures $E_e$, $V_L$, $V_1$, V, $C$, $T$ and $T'$\nllabel{host-process-01}\\
    $C=\emptyset$\nllabel{host-process-02}\\
    
    \For{$i = 1, 2, \ldots, |V| - 3$ \nllabel{host-for}\nllabel{host-process-03}}
    {
       Launch $|SM| \cdot MaxSMSize$ threads each one running Algorithm~\ref{alg:cc_visit_parallel}\\\nllabel{host-process-04}
       $T \leftarrow T'$\nllabel{host-process-05}\\
       Wait all threads to finish\nllabel{host-process-06}
    }
    Return $C$\nllabel{host-process-07}\\
   \caption{$HostProcess(G, \ell)$
   \label{alg:hostprocess}}
  \end{algorithm2e}
 \end{minipage}
\end{center}

\section{Computational Experiments}
\label{sec:results}

Both parallel and sequential~\footnote{We emphasize that the sequential algorithm is the fastest one known up to now, as described in~\cite{DCLJ2014}.} algorithms were coded in the C++ language and compiled using a GNU compiler (g++ version 4.8.2 with parameters ``\verb|-O3 -mcmodel=medium -m64 -g -W -Wall|''). The parallel algorithm also used OpenCL 1.2 with the AMD Software Development Kit 2.9.1. All experiments were performed on a computer with an AMD FX-9590 Black Edition Octa Core CPU, with clock ranging from 4.7GHz to 5.0GHz, 32GB of RAM, runnning Ubuntu 14.04 64-bits operating system. The computer had a Radeon SAPPHIRE R9 290X Tri-X OC GPU video card, with 4GB of memory. The architecture of such a video card provides 2816 stream processing units and an enhanced engine clock of up to 1040Mhz. Its memory is clocked at 1300MHz (5.2GHz effectively).


In order to evaluate the benefits of the parallel algorithm over the sequential one, in terms of processing time to enumerate all chordless cycles, we performed intensive experiments with several datasets. We used twenty three graphs for the experiments, joined in three groups. The first group consists of ten graphs presented in well known databases of ecological studies~\cite{C1978}. These graphs, which have already been considered by Sokhn et al.~\cite{SBBHN2013}, are formed by directed edges and represent {\it food webs}. For the application of such graphs in the current experiments, it was necessary to transform them into undirected {\it niche overlap} graphs. This was done using the definitions provided by Wilson and Watkins~\cite{WW1990}. 

The second group consists of three urban traffic networks, regarding the cities of Sioux Falls, ??? and part of the downtown area of the city of Goi\^ania, the capital of the state of Goi\'as, in Brazil (illustrated in Figure~\ref{fig:Goiania_Downtown}). Finally, the last group contains well structured graphs representing a cycle, a wheel, some bipartite graphs and grids.


Table~\ref{tab:Tempo_Exec1} presents details of each graph. It shows the name of the graph, the numbers $n$ and $m$ of vertices and edges respectively, and the maximum degree $\Delta$. The remaining columns contain information produced by our algorithms. Column $C_3$   gives the number of cycles of length three. They are found at the first stage of the sequential and the parallel algorithms. Column $\#clc$ provides the number of chordless cycles with length greater than three, found in the graph. 

The sequential and parallel algorithms were run ten times for each graph. The average running times of the ten executions are presented in the table in milliseconds. Column $T_{seq}$ are the processing times of the sequential algorithm. The next two columns are the average times related to the parallel GPU algorithm. The first column ($T_{par-proc}$) contains only the processing time spent by the GPU kernels at the first and second stages, plus the time for the sequential degree labeling preprocessing; the second column ($T_{par-total}$) has the total time of the parallel code; this includes the processing time ($T_{par-proc}$) plus the communication time between the host and the GPU in order to transfer the graph structure and the solution set $C$. The last column of Table~\ref{tab:Tempo_Exec1} is the speedup of the parallel algorithm over the sequential algorithm (given by $T_{par-total}/T_{seq}$).

\begin{table}[htpb]
\begin{minipage}[h]{1.\linewidth}
\centering\small
\caption{Running time to enumerate all chordless cycles on niche overlap graphs and on other well known graphs.}
\label{tab:Tempo_Exec1}
\begin{tabular}{lrrrrrrrrr}
\hline
\textbf{Name}	& $\boldsymbol{n}$ 	& $\boldsymbol{m}$ 	& $\boldsymbol{\Delta}$ 	& $\boldsymbol{C_3}$		& \textbf{\#clc}& $\boldsymbol{T_{seq}}$	& $\boldsymbol{T_{par-proc}}$& $\boldsymbol{T_{par-total}}$& \textbf{\emph{Speedup}}	\\
\hline
CrystalD	& 24	& 86		& 14		& 293		& 0		& 0.333		& 0.182		& 0.622		& 0.536		 \\
ChesUpper	& 37	& 85		& 15		& 167		& 0		& 0.370		& 0.160		& 0.656		& 0.564		 \\
Narragan	& 35	& 168		& 22		& 586		& 0		& 0.548		& 0.197		& 0.709		& 0.773		 \\
Chesapeake	& 39	& 90		& 11		& 157		& 0		& 0.150		& 0.188		& 0.700		& 0.214		 \\
Michigan	& 39	& 175		& 27		& 587		& 0		& 0.614		& 0.197		& 0.698		& 0.879		 \\
Mondego	& 46	& 206		& 24		& 886		& 0		& 0.725		& 0.207		& 0.773		& 0.938		 \\
Cypwet	& 71	& 842		& 46		& 8946		& 0		& 6.417		& 0.258		& 0.892		& 7.196		 \\
Everglades	& 69	& 1214		& 56		& 15627		& 710		& 12.407	& 0.388		& 1.478		& 8.395		 \\
Mangrovedry	& 97	& 2132		& 80		& 30659		& 27426		& 102.475	& 1.822		& 6.510		& 15.741	 \\
Floridabay	& 128	& 3249		& 98		& 62389		& 85976		& 366.495	& 2.518		& 15.095	& 24.279	 \\
\hline
Goi\^ania	& 43	& 75		& 4		& 5		& 9311		& 39.594	& 0.216		& 3.081		& 12.849	 \\
SiouxFalls	& 24	& 76		& 5		& 2		& 176		&  1.339	& 1.138		& 1.812		& 0.739	 \\
???	& ???	& ???		& ???		& ???		& ???		&  ???	& ???		& ???		& ???	 \\ \hline
$C_{100}$		& 100	& 100		& 2		& 0		& 1		& 0.149		& 0.165		& 0.770		& 0.193		 \\
Wheel 100	& 101	& 200		& 100		& 100		& 1		& 0.225		& 0.778		& 1.229		& 0.183		 \\
$K_{8,8}$	& 16	& 64		& 8		& 0		& 784		& 0.473		& 0.197		& 0.599		& 0.790		 \\
$K_{50,50}$	& 100	& 2500		& 50		& 0		& 1500625	& 600.661	& 4.867		& 10.391	& 57.805	 \\
Grid $4{\times}10$	& 40	& 66		& 4		& 0		& 1823		& 15.430	& 0.185		& 1.993		& 7.742		 \\
Grid $5{\times}6$	& 30	& 49		& 4		& 0		& 749		& 2.610		& 0.167		& 1.249		& 2.090		 \\
Grid $5{\times}10$	& 50	& 85		& 4		& 0		& 52620		& 199.132	& 1.982		& 12.718	& 15.658	 \\
Grid $6{\times}6$ 	& 36	& 60		& 4		& 0		& 3436		& 7.889		& 0.203		& 1.570		& 5.025		 \\
Grid $6{\times}10$	& 60	& 104		& 4		& 0		& 800139	& 2906.009	& 6.284		& 18.989	& 153.034	 \\
Grid $7{\times}10$	& 70	& 123		& 4		& 0		& 8136453	& 36955.470	& 54.840	& 286.212	& 129.119	 \\
Grid $8{\times}10$\footnote{Due to high memory consumption for storing set $T$ when processing Grid $8{\times}10$, both the sequential and parallel algorithms were modified to not store the chordless cycles, but only to count them.}	& 80	& 142		& 4		& 0		& 71535910	& 427091.02	& 1655.147	& 8697.081	& 49.107	 \\
\hline
\end{tabular}
\end{minipage}
\end{table}

\subsection{Analysis of the results}
\label{subsec:analysis}

The benefits of the parallel algorithm over the sequential one depend on the nature of the graph. As we can see, the gain in speedup is, in general, proportional to the number of chordless cycles and paths, with speedups ranging from 12$\times$ to 153$\times$ for the most complex cases (with $|C|\geq 100.000$). When the graph had not many chordless cycles and paths, the sequential algorithm overcame the parallel GPU code. 

Note however that, almost all worst cases (when the speedup was less than 1, indicating a better performance to the sequential algorithm),  the most expensive activity in the parallel algorithm was the data communication between the host and the GPU device. So, when considered only the GPU kernel time (column $T_{par-proc}$), the parallel algorithm is very competitive. Furthermore, the parallel algorithm executed in less than 0.002 seconds for all non-competitive cases.

\begin{figure}[ht]
 \begin{tabular} {cc} 
  \parbox[c]{3.1in}{\includegraphics[width=3.1in]{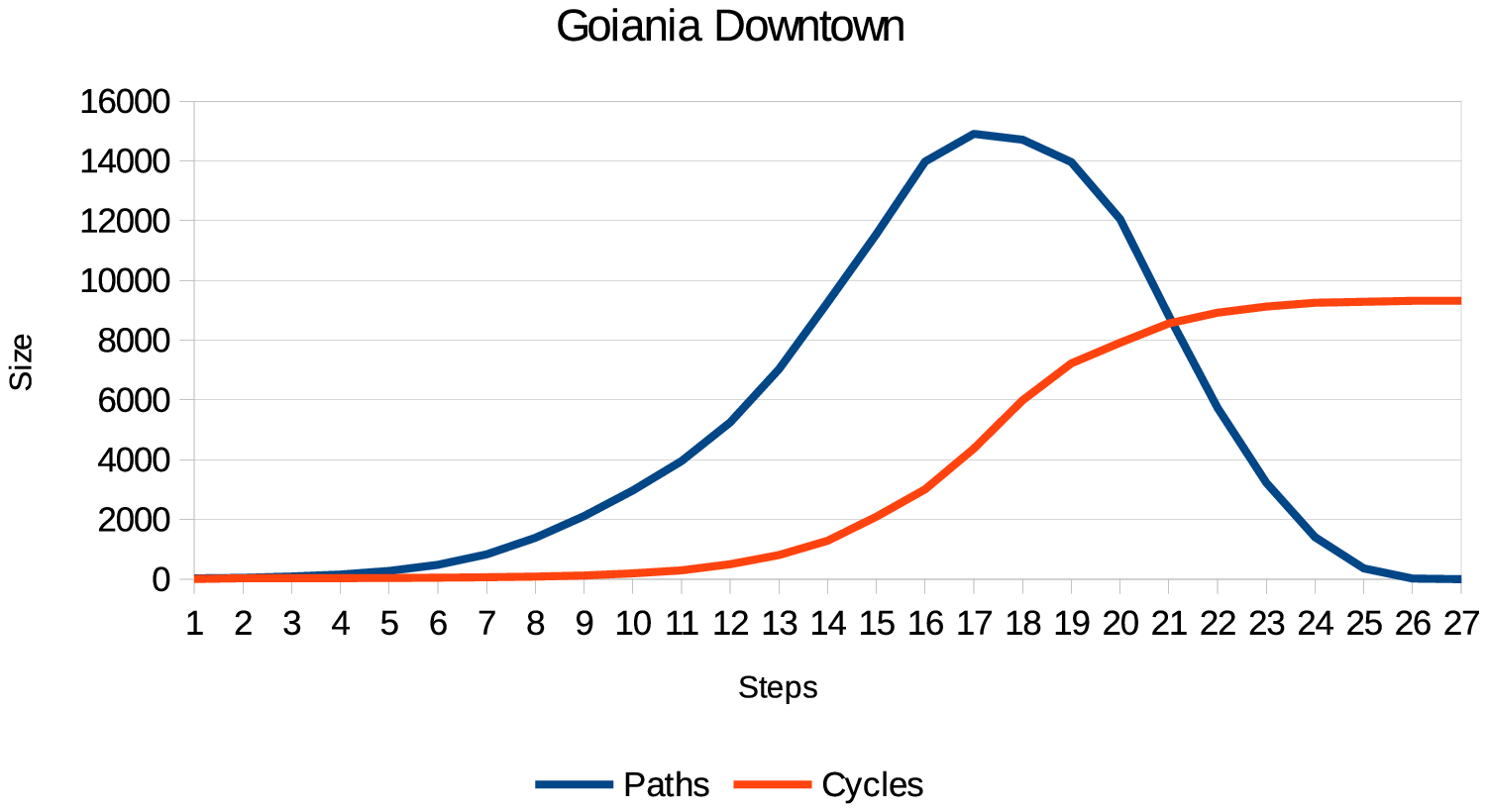}} & \parbox[c]{3.1in}{\includegraphics[width=3.1in]{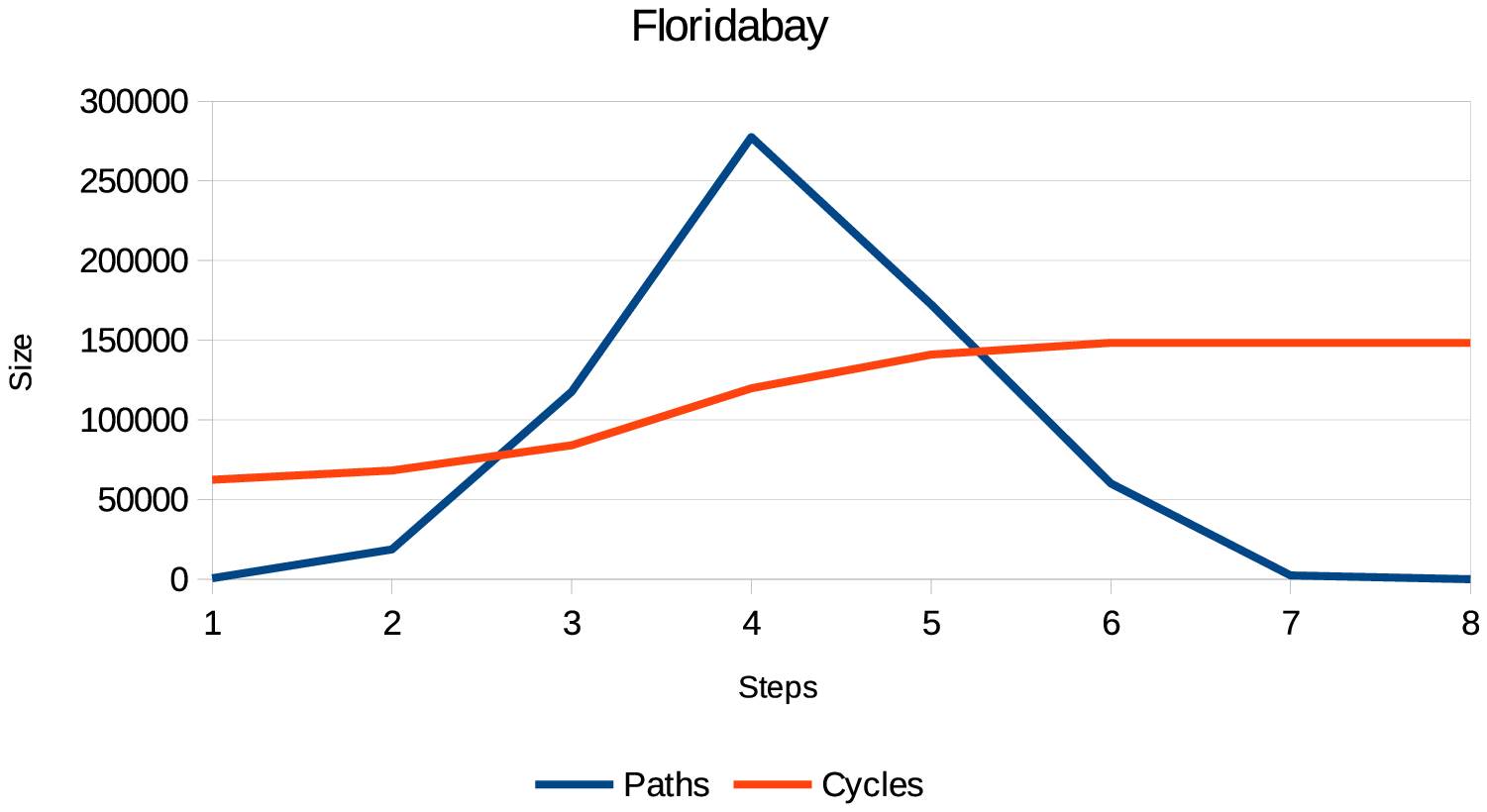}} \\
  \parbox[c]{3.1in}{\includegraphics[width=3.1in]{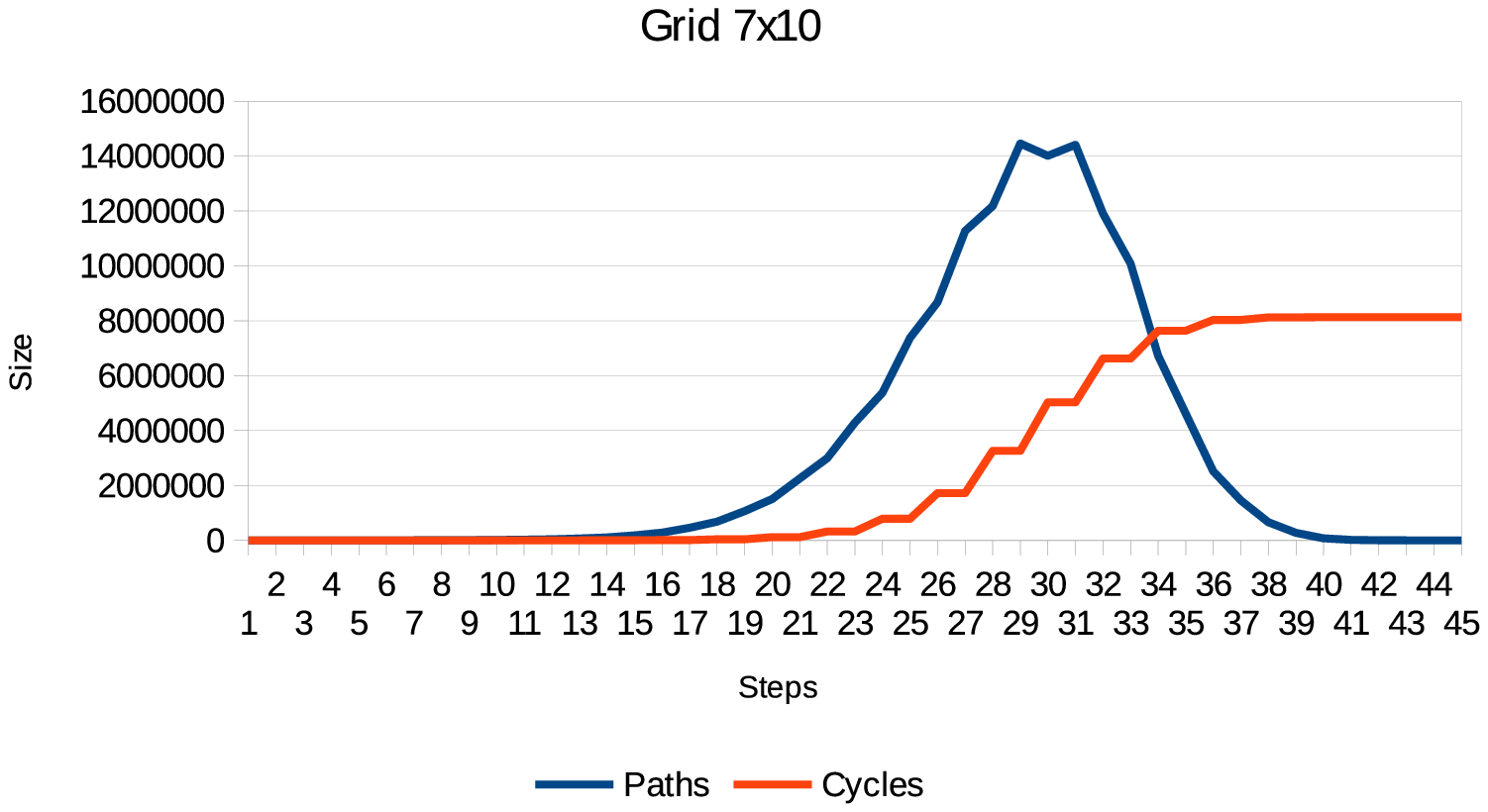}} & \parbox[c]{3.1in}{\includegraphics[width=3.1in]{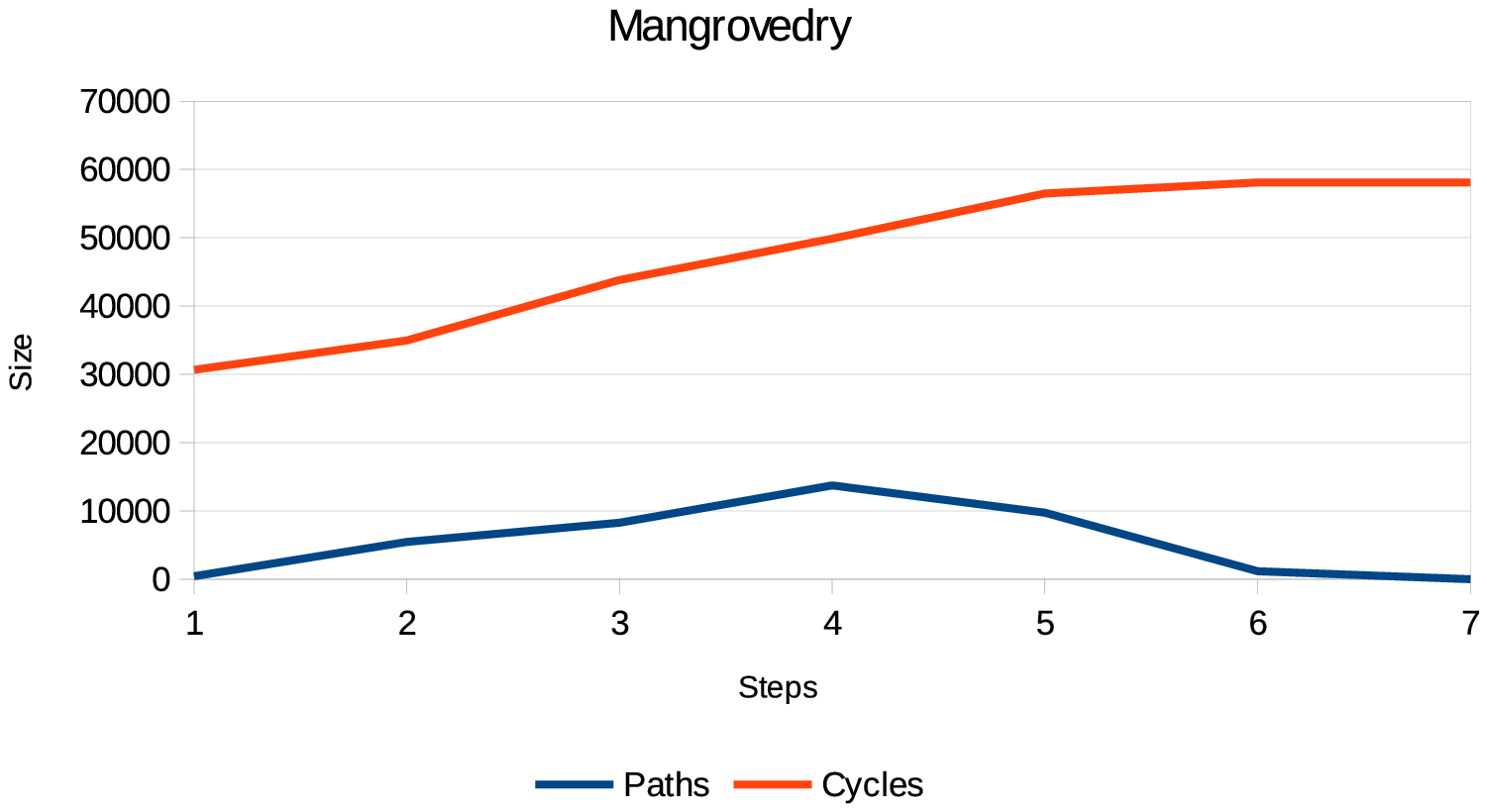}} \\
 \end{tabular}
 \caption{Sizes of T and C for four graphs.}
 \label{tab:t_c_sizes}
\end{figure}

It is useful to see, as well,  the evolution of sets $C$ and $T$ in size during an execution of the two stages of the parallel algorithm. This gives a hint about the amount of computation done by the parallel threads over time, and how much synchronization was necessary for writing on the data structures that hold such sets. Figure~\ref{tab:t_c_sizes} shows this evolution for the graphs Floridabay, Mangrovedry, Grid $7{\times}10$ and Goi\^ania. The blue (darker) line in each chart represents the size of set $T$ at each call of the kernels; the red (lighter) line shows the change on the size of set $C$. The X axis represents the results of both stages and also implies the size of all paths in the current set $T$. Step 1 in the chart represents the result of the first stage of our algorithm. The following steps are related to the output of each iteration (kernel call) of the second stage.  

At the beginning of the computation, both $C$ and $T$ sets are empty. Then they are initialized by the first stage of the parallel algorithm. As the algorithm processes through the second stage, new chordless paths are created by extending smaller paths with adjacent vertices. This causes the set $T$ to increase in size. In this case, more synchronization for writing in $T$ and $C$ occurs. Latter, the extension of some paths result in chordless cycles (that are then added to $C$) or in cycles with chords (that are just discarded). The overall process results in a wave shape for the evolution chart of $T$ and a more soft increasing curve for the evolution chart of $C$.

It is interesting to note that even with a very high peak of the size of $T$ for the graph Grid $7{\times}10$, with 14 millions of chordless paths stored, the performance of the parallel algorithm was much superior than that of the sequential one (with a speedup of $\approx 129{\times}$).

A curious case was graph Mangrovedry. Many chordless cycles of size three (around 30.000) were found right at the first stage of the parallel algorithm. The second stage of the algorithm performed only seven steps (similarly to graph Floridabay), which doubles the size of $C$ but with chordless cycles of length at most 9 (recall that the initial chordless paths have length 3, as found by the first stage of the algorithm, and they grow one edge at every iteration of the second stage).

\section{Conclusions}
\label{sec:conclusions}
In this paper, was presented a parallel algorithm for GPUs to enumerate all chordless cycles of a given undirected graph. The algorithm is based on a previous work done by some of the authors, which resulted in an already fast sequential algorithm for the same problem. The parallel algorithm works in two stages: in the first stage it computes an initial set of triplets and an initial set of chordless paths for expansion; in the second stage, all chordless paths are analyzed and then expanded or removed. The parallel algorithm takes advantage of the GPU architecture by distributing many tasks that are necessary in each stage to the groups of GPU processing units. A compact data structure for graph representation, distinct types of memories and the persistent thread technique were employed for allowing a more efficient usage of the GPU memory and processing power. 

Experiments were done with several graphs and they showed that the benefits of the parallel algorithm depends on a large number of the chordless cycles and chordless paths in the input graph. For the graphs with more than 100.000 chordless cycles or paths, the speedup of the parallel algorithm over the sequential one was from $\approx 12$ to $153$ times. The cases for which the parallel algorithm was worse (took longer than the sequential algorithm) were the ones with very few chordless cycles and most of the exceeding time was spent with data transfer between the CPU and the GPU. For those base cases, our implementation still took less than 0.002 seconds to find all chordless cycles.

Up to our knowledge, this is the first parallel GPU-based algorithm for the problem of enumerating all chordless cycles. Note, however, that memory size on a GPU is still a restricting factor since the data structures cannot be larger than the maximum supported texture size. Such hardware constraints limit the size of the problems and solutions that can be dealt with by the GPUs. Thus, as a future work, we are planning to develop a new data transportation protocol between the ordinary RAM memory and the GPU memory in order to open space when necessary and allow to enumerate chordless cycles for much larger datasets. We are also implementing a parallel algorithm for computing the degree labeling. Deleting a vertex during such a computation can lead to a major change in the graph (the decrease of one unit of the degree of every adjacent vertex), what indicates that the labeling process has an inherent sequential nature. However, one could update the degree of all vertices in parallel in constant time using $n \cdot \Delta$ processors. Then, the smallest degree can be found through a parallel reduction in time $\mathcal{O}(\log (n))$ with $n$ threads. Repeating this process $n-1$ times provides the desirable result with total time  
$\mathcal{O}(n\log (n))$.

\section*{Acknowledgement}

We thank the Brazilian research supporting agencies CAPES (Coordena\c{c}\~{a}o de Aperfei\c{c}oamento de Pessoal de N\'{i}vel Superior) and FAPEG (Funda\c{c}\~{a}o de Amparo \`{a} Pesquisa do Estado de Goi\'{a}s) for providing Ph.D. scholarships to the first and the second authors, respectively.

\end{document}